\documentclass[runningheads]{llncs}

\usepackage{amsmath, amsfonts, amssymb}

\usepackage{graphicx}
\usepackage{url}
\usepackage{blindtext}
\usepackage{enumitem}
\usepackage{graphicx}
\usepackage{amssymb}
\usepackage{amsmath}
\usepackage{algpseudocode}
\usepackage{caption}
\usepackage{subfig}
\usepackage{cite}
\usepackage{hhline}
\usepackage{authblk}
\usepackage{comment}
\usepackage[export]{adjustbox}
\usepackage{ulem}
\usepackage{makecell}

\usepackage{tikz}

\begin{document}

\mainmatter  

\title{3D-StyleGAN: A Style-Based Generative Adversarial Network for Generative Modeling of Three-Dimensional Medical Images}

\author{Sungmin Hong$^1$, Razvan Marinescu$^2$, Adrian V. Dalca$^{2,3}$, Anna K. Bonkhoff$^1$, Martin Bretzner$^{1,4}$, Natalia S. Rost$^1$ and Polina Golland$^2$}

\institute{$^1$ JPK Stroke Research Center, Department of Neurology, Massachusetts General Hospital, Harvard Medical School, Boston, MA, USA.\\
$^2$ Computer Science and Artificial Intelligence Laboratory, Massachusetts Institute of Technology, Cambridge, MA, USA. \\  
$^3$ A.A. Martinos Center for Biomedical Imaging, Department of Radiology, Massachusetts General Hospital, Harvard Medical School, Cambridge, MA, USA.\\
$^4$ Univ.  Lille, Inserm, CHU Lille, U1172 - LilNCog (JPARC) - Lille Neurosciences \& Cognition, F-59000 Lille, France.
}

\index{Hong, Sungmin} 
\index{Marinescu, Razvan} 
\index{Dalca, Adrian}
\index{Bonkhoff, Anna}
\index{Bretzner, Martin}
\index{Rost, Natalia} 
\index{Golland, Polina} 

\authorrunning{S. Hong et al.}
\titlerunning{3D-StyleGAN for Generative Modeling of 3D Medical Images}
\maketitle
\setcounter{footnote}{0}

\begin{abstract}
Image synthesis via Generative Adversarial Networks (GANs) of three-dimensional (3D) medical images has great potential that can be extended to many medical applications, such as, image enhancement and disease progression modeling.
However, current GAN technologies for 3D medical image synthesis need to be significantly improved to be readily adapted to real-world medical problems.
In this paper, we extend the state-of-the-art StyleGAN2 model, which natively works with two-dimensional images, to enable 3D image synthesis. In addition to the image synthesis, we investigate the controllability and interpretability of the 3D-StyleGAN via style vectors inherited form the original StyleGAN2 that are highly suitable for medical applications: (i) the latent space projection and reconstruction of unseen real images, and (ii) style mixing. We demonstrate the 3D-StyleGAN's performance and feasibility with $\sim$12,000 three-dimensional full brain MR T1 images, although it can be applied to any 3D volumetric images. Furthermore, we explore different configurations of hyperparameters to investigate potential improvement of the image synthesis with larger networks. The codes and pre-trained networks are available online: \textit{https://github.com/sh4174/3DStyleGAN}.
\end{abstract}


\section{Introduction}
\label{sec:Intro}

Generative modeling via Generative Adversarial Networks (GAN) has achieved remarkable improvements with respect to the quality of generated images~\cite{goodfellow2014generative,arjovsky2017wasserstein,brock2018large,zhu2017unpaired,karras2020analyzing}. StyleGAN2, a style-based generative adversarial network, has been recently proposed for synthesizing highly realistic and diverse natural images. It progressively accounts for multi-resolution information of images during training, and controls image synthesis using style vectors that are fed at each block of a style-based generator network~\cite{karras2017progressive,karras2019style,karras2020analyzing,karras2020training}. 
It showed the outstanding quality of generated images and the enhanced control and interpretability on image synthesis compared to previous generative models~\cite{brock2018large,arjovsky2017wasserstein,karras2019style,karras2020analyzing}.  
Although it has great potential for medical applications due to its enhanced performance and controllability, it has not been extended to the generative modeling of 3D medical images in our best knowledge. 

One of the challenges of using generative models such as GANs for medical applications is that medical images are often three-dimensional (3D) and have a significantly higher number of voxels compared to two-dimensional natural images.
Due to the high-memory requirements of GANs, it is often not feasible to directly apply large networks for 3D image synthesis~\cite{brock2018large,kwon2019generation,volokitin2020modelling}. 
To address this issue, some models for 3D image synthesis generate the image in successive 2D slices, which are then combined to render a 3D image while accounting for slice-wise relationship~\cite{volokitin2020modelling}. However, these methods often result in discontinuity between slices, and the interpretation and manipulation of the slice-specific latent vectors is complicated in this setting ~\cite{volokitin2020modelling}.


In this paper, we present 3D-StyleGAN to enable synthesis of high-quality 3D medical images by extending the StyleGAN2. 
We made several changes to the original StyleGAN2 architecture: (1) we replaced 2D operations, layers, and noise inputs with 3D, and (2) significantly decreased the depths of filter maps and latent vector sizes. We trained different configurations of 3D-StyleGAN on a collection of $\sim$12,000 T1 MRI brain scans from ~\cite{dalca2018anatomical}. We additionally make the following contributions: (1) we show the possibility of synthesizing realistic 3D brain MRIs at 2mm resolution, corresponding to 80x96x112 voxels, (2) we show that projection of unseen test images to the latent space can be achieved and results in reconstructions of high fidelity to the input images by a projection function suitable for medical images and (3) we demonstrate how StyleGAN2's style-mixing can be used to ``transfer'' anatomical variation across images. We discuss the performance and feasibility of the 3D-StyleGAN with limited filter depths and latent vector sizes for 1mm isotropic resolution full brain T1 MR images. The source code and pre-trained networks are publicly available in \textit{https://github.com/sh4174/3DStyleGAN}.







\section{Methods}
\label{Methods}

\begin{figure}[htb!]
\centering
  \subfloat{ \includegraphics[width=120mm]{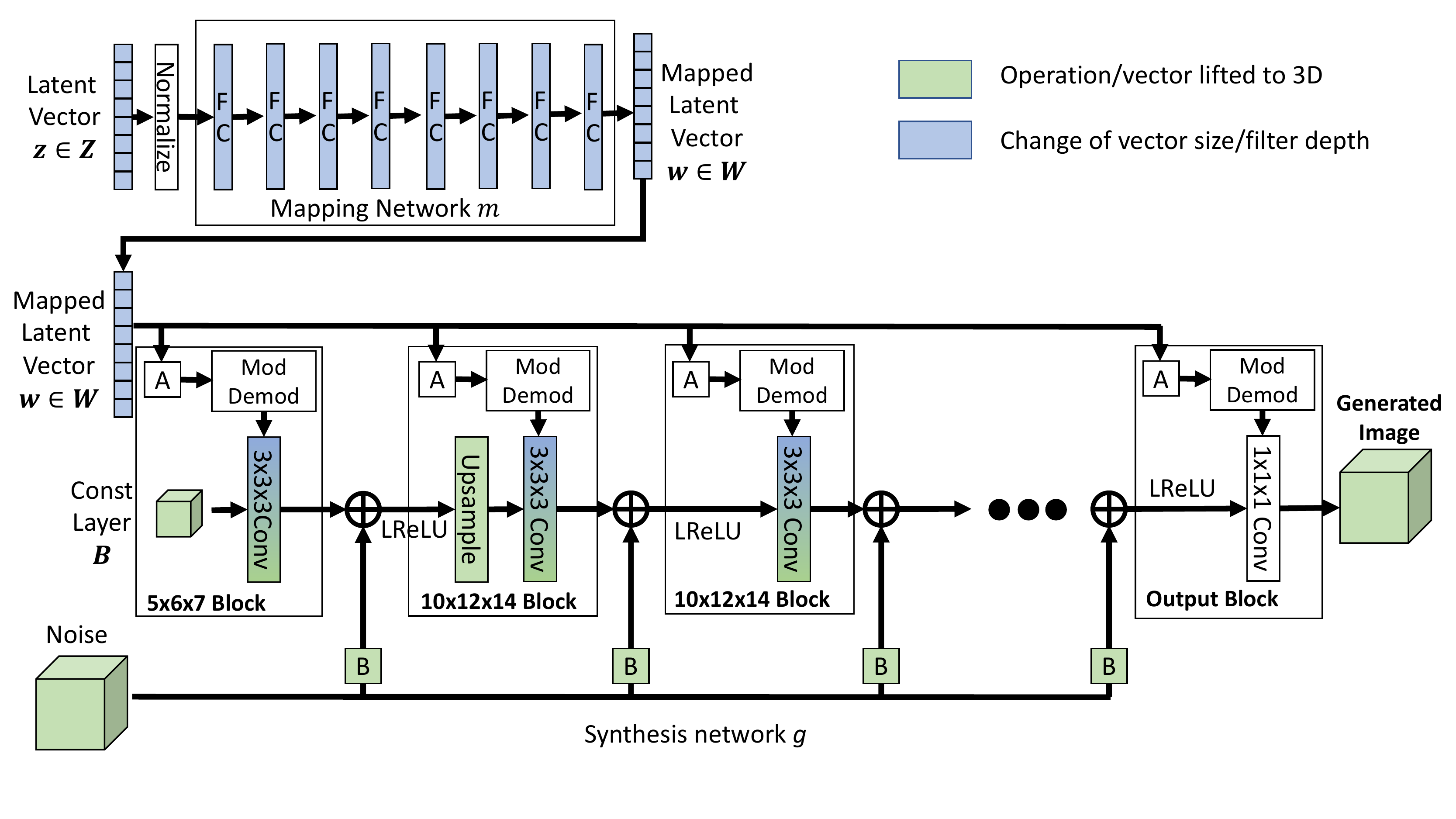} } 
\caption{The style-based generator architecture of 3D-StyleGAN. The mapping network $m$, made of 8 fully-convolutional layers, maps a normalized latent vector $\mathbf{z}$ to an intermediate latent vector $\mathbf{w}$. This is then transformed by a learned feature-wise affine transform $A$ at each layer, then further used by modulation (Mod) and demodulation (Demod) operations that are applied on the weights of each convolution layer ~\cite{huang2017arbitrary,dumoulin2018feature}. The synthesis network starts with a constant 5x6x7 block and applies successive convolutions with modulated weights, up-sampling, and activations with Leaky ReLU. 3D noise maps, downscaled to the corresponding layer resolutions, are added before each nonlinear activation by LReLU. At the final layer, the output is convolved by a  1$\times$1$\times$1 convolution to generate the final image. $\mathbf{B}$ is the size of base layer (5$\times$6$\times$7). The discriminator network (3D-ResNet) was omitted due to space constraint.}
\label{fig:GenArch}
\end{figure}

Figure~\ref{fig:GenArch} illustrates the architecture of the 3D-StyleGAN. A style-based generator was first suggested in~\cite{karras2019style} and updated in~\cite{karras2020analyzing}.
We will briefly introduce the style-based generator suggested in~\cite{karras2020analyzing} and then summarize the changes we made for enabling 3D image synthesis.

\noindent \textbf{StyleGAN2:}    
A latent vector $\mathbf{z} \in \mathcal{Z}$ is first normalized and then mapped by a mapping network $m$ to $\mathbf{w} = m(\mathbf{z})$, $\mathbf{w} \in \mathcal{W}$. 
A synthesis network $g$ starts with a constant layer with a base size $\mathbf{B} \in \mathbb{R}^d$, where 
$d$ is the input dimension (two for natural images).
The transformed $A(\mathbf{w})$ is modulated ($Mod$) with trainable convolution weights $w$ and then demodulated ($Demod$), which act as instance normalization to reduce artifacts caused by arbitrary amplification of certain feature maps.
Noise inputs scaled by a trainable factor $B$ and a bias $b$ are added to the convolution output at each style block.  
After the noise addition, the Leaky Rectifier Linear Unit (LReLU) is applied as nonlinear activation~\cite{maas2013rectifier}. 
At the final layer in the highest resolution block, the output is fed to the 1$\times$1 convolution filter to generate an image. 

\noindent \textbf{Loss Functions and Optimization: } For the generator loss function, StyleGAN2 uses the logistic loss function with the path length regularization~\cite{karras2020analyzing}:
\begin{equation}
\mathbb{E}_{\mathbf{w}, \mathbf{y} \sim \mathcal{N}(0, \mathbf{I})} ( || \nabla_\mathbf{w} ( g(\mathbf{w}) \cdot \mathbf{y}) ||_2 - a )^2,
\end{equation}
where $\mathbf{w}$ is a mapped latent vector, $g$ is a generator, $\mathbf{y}$ is a random image following a normal intensity distribution with the identity covariance matrix, and $a$ is the dynamic constant learned as the running exponential average of the first term over iterations~\cite{karras2020analyzing}.
It regularizes the gradient magnitude of a generated image $g(\mathbf{w})$ projected on $\mathbf{y}$ to be similar to the running exponential average to make the (mapped) latent space $\mathcal{W}$ smoother. 
The generator regularization was applied once in every 16 minibatches following the lazy regularization strategy. 
For the discriminator loss function, StyleGAN2 uses the standard logistic loss function with $\mathbf{R}_{1}$ or $\mathbf{R}_2$ regularizations. 

\subsection{3D-StyleGAN} We modified StyleGAN2 for 3D image synthesis by replacing the 2D convolutions, upsampling and downsampling with 3D operations. We started from StyleGAN2 configuration F (see \cite{karras2020analyzing}), but switched back to the original StyleGAN generator with $(De)Mod$ operators, as it showed best performances in preliminary results for 3D images~\cite{karras2020analyzing, he2016deep,hara3dcnns}. We used the 3D residual network for the discriminator. We used the standard logistic loss function without regularization for the discriminator loss function that showed the best empirical results.

\noindent \textbf{Image Projection: }An image (either generated or real) can be projected to the latent space by finding $\mathbf{w}$ and stochastic noise $\mathbf{n}$ at each layer that minimize a distance between an input image $I$ and a generated image $g(\mathbf{w}, \mathbf{n})$. 
In the original StyleGAN2, LPIPS distance was used~\cite{zhang2018unreasonable}. 
However, the LPIPS distance uses the VGG16 network trained with 2D natural images, which is not straightforwardly applicable to 3D images~\cite{simonyan2014very}. 
Instead, we used two mean squared error (MSE) losses, one computed at full resolution, and the other at an (x8) downsampled resolution. The second loss was added for stability, and to ensure that the optimization does not converge to a local minimum.  



\noindent \textbf{Configurations: } We tested different resolutions, filter depths, latent vector sizes, minibatch sizes, which are all summarized in Table~\ref{tab:config}. Because of the high-dimensionality of 3D images, the filter map depths of the 3D-StyleGAN needed to be significantly lower than in the 2D version.
We tested five different feature depths: 16, 32, 64, and 96, with 2mm isotropic resolution brain MR images to investigate how different filter depths would affect the quality of generated images. 
For 1mm isotropic images, we used the filter depth of 16 for the generator and discriminator, 32 filter depth for the mapping network, and 32 latent vector size that our computational resource allowed. 

\noindent \textbf{Slice-wise Fr\'{e}chet Inception Distance Metric: }  Conventionally, the Fr\'{e}chet Inception Distance (FID) score is measured by comparing the distributions of randomly sampled real images from a training set and generated images~\cite{heusel2017gans}. 
Because FID relies on the Inception V3 network that was pre-trained with 2D natural images, it is not directly applicable to 3D images~\cite{simonyan2014very}. 
Instead, we measured the Fr\'{e}chet Inception Distance (FID) scores on the middle slices on axial, coronal, and sagittal planes of the generated 3D images~\cite{heusel2017gans}.

\noindent \textbf{Additional Evaluation Metrics: } In addition to the slice-wise FID metric, we evaluated the quality of generated images with the batch-wise squared Maximum Mean Discrepancy (bMMD$^2$) as suggested in~\cite{kwon2019generation} and~\cite{volokitin2020modelling}. Briefly, MMD$^2$ measures the distance between two distributions with finite sample estimates with kernel functions in the reproducing kernel Hilbert space~\cite{gretton2012kernel}. The bMMD$^2$ measures the discrepancy between images in a batch with a dot product as a kernel~\cite{kwon2019generation,volokitin2020modelling}.
To measure the diversity of generated images, we used the pair-wise Multi-Scale Structural Similarity (MS-SSIM)~\cite{kwon2019generation,volokitin2020modelling}. 
It measures the perceptual diversity of generated images by measuring the mean of the MS-SSIM scores of pairs of generated image~\cite{odena2017conditional}.





\begin{table}[tb!]
\centering
\caption{The list of configurations of 3D-StyleGAN with different filter map depths, latent vector sizes and minibatch sizes.}
\begin{tabular}{|l|l|l|l|l|}
\hline
Config.  & Filter Depth & Latent Vector Size & \begin{tabular}[c]{@{}l@{}}Minibatch Size \\ \{B, 2B, $2^2$B, $2^3$B, $2^4$B\}\end{tabular}                                    & \begin{tabular}[c]{@{}l@{}}\# trainable \\ params\end{tabular} \\ \hline
2mm-fd96 & 96           & 96                 & \{32, 32, 16, 16\}            & 6.6M                                                           \\ \hline
2mm-fd64 & 64           & 64                 & \{32, 32, 16, 16\}             & 3.0M                                                           \\ \hline
2mm-fd32 & 32           & 128                & \{32, 32, 16, 16\}             & 0.9M                                                           \\ \hline
2mm-fd16 & 16           & 64                 & \{32, 32, 16, 16\}             & 0.2M                                                           \\ \hline
1mm-fd16 & 16           & 64                 & \{64, 64, 32, 16, 16\} & 0.2M                                                           \\ \hline
\end{tabular}
\label{tab:config}
\end{table}

\section{Results}
\label{sec:Results}

For all experiments, we used 11,820 brain MR T1 images from multiple publicly available datasets: ADNI, OASIS, ABIDE, ADHD2000, MCIC, PPMI, HABS, and Harvard GSP~\cite{mueller2005ways,marcus2007open,di2014autism,milham2012adhd,gollub2013mcic,marek2011parkinson,dagley2017harvard,holmes2015brain,dalca2018anatomical}. All images were skull-stripped, affine-aligned, resampled to 1mm isotropic resolution, and trimmed to 160$\times$192$\times$224 size using FreeSurfer~\cite{dalca2018anatomical,fischl2012freesurfer}.
For the experiments with 2mm isotropic resolution, the images were resampled to an image size of 80$\times$96$\times$112.
Among those images, 7,392 images were used for training. The rest of 4,329 images were used for evaluation. The base layer was set to $\mathbf{B}$=$5\times$6$\times$7 to account for the input image size.  
For each experiment, we used four NVIDIA Titan Xp GPUs for the training with 2mm-resolution images, and eight GPUs for the training with 1mm-resolution images. The codes were implemented with Tensorflow 1.15 and Python 3.6~\cite{tensorflow2015-whitepaper}.

\newcommand{\incw}[2]{\includegraphics[height=#2]{#1}}

\newcommand{\pp}{projected/image0001}
\newcommand{\pt}{projected/image0002}

\begin{figure}[tb!]
\centering
  \subfloat{ \includegraphics[width=120mm]{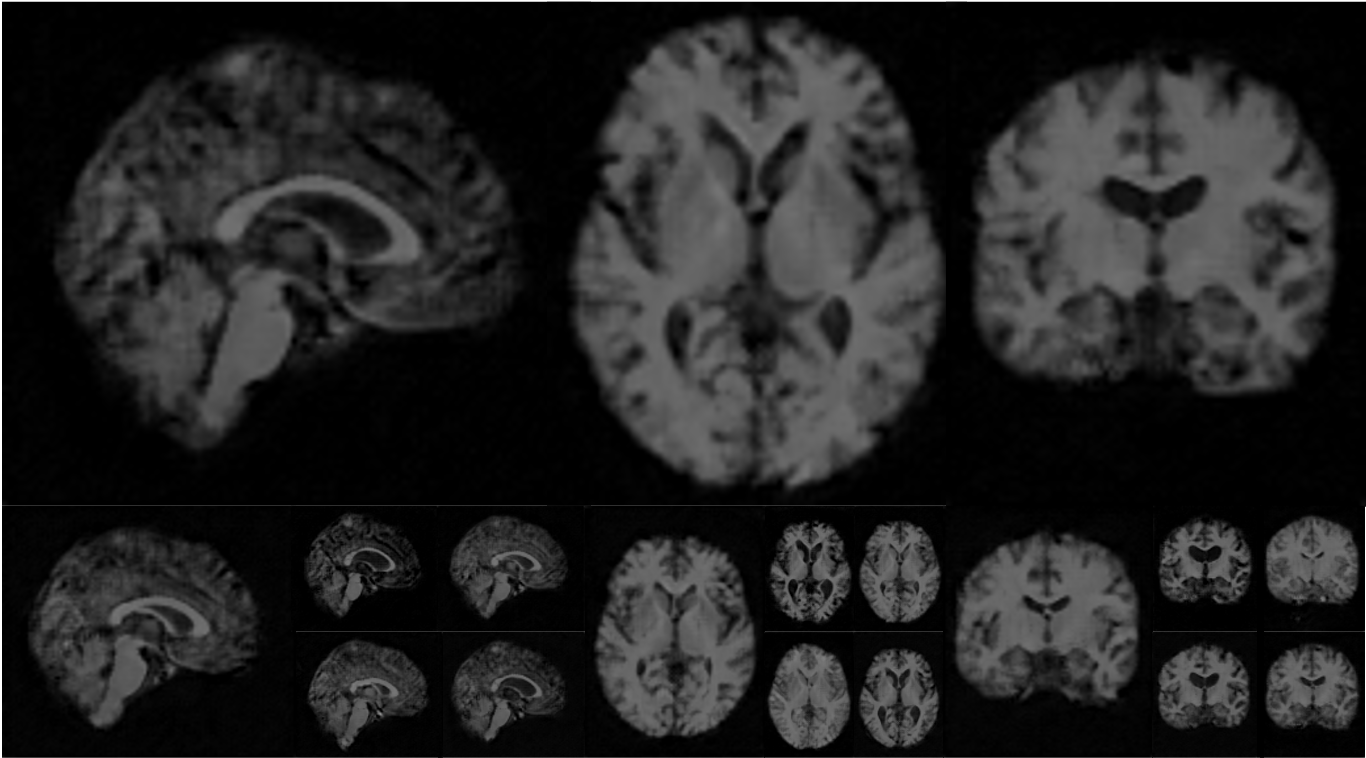} } 
\caption{Uncurated randomly generated 3D images by the 3D-StyleGAN. The images are generated by the network trained with configuration 2mm-fd96. The middle slices of sagittal, axial, and coronal axes of five 3D images were presented.}
\label{fig:GenImg_iso}
\end{figure}

\begin{figure}[htb!]
\centering
  \subfloat[2mm-fd96]{ \includegraphics[width=60mm,height=30mm]{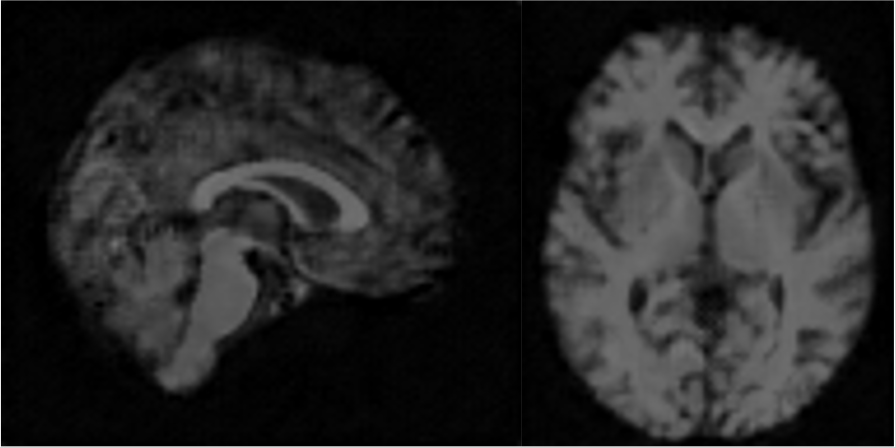} }
  \subfloat[2mm-fd64]{ \includegraphics[width=60mm,height=30mm]{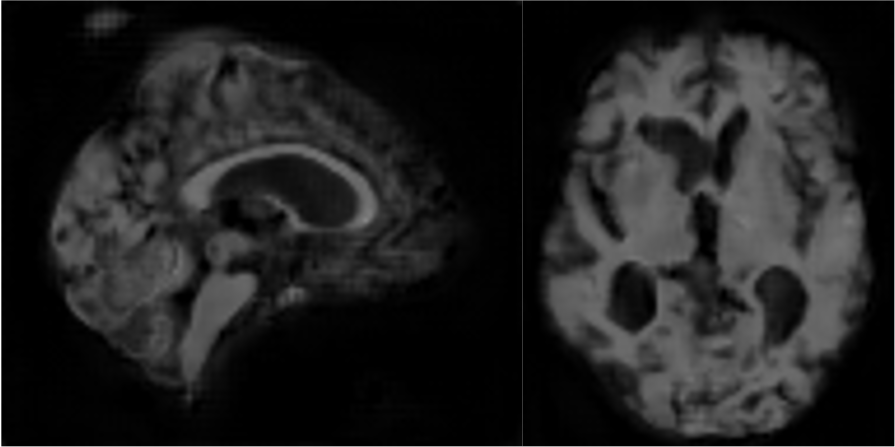} }
\caption{The qualitative comparison between the networks trained with the filter map depths of (a) 96 (2mm-fd96) and (b) 64 (2mm-fd64). While the generated images of the 2mm-fd64 are sharper, e.g., cerebellum and sulci, the overall brain structures are not as coherent as those in the images generated with 2mm-fd96.}
\label{fig:96_64_comp}
\end{figure}

\begin{figure}[htb!]
\centering
\begin{tikzpicture}
    
    \node[inner sep=0pt] (target) at (0,1)
    {\incw{\pp-target}{2cm}\incw{\pp-target_y}{2cm}\incw{\pp-target_z}{2cm}~\incw{\pt-target}{2cm}\incw{\pt-target_y}{2cm}\incw{\pt-target_z}{2cm}};
    \node[inner sep=0pt] (target) at (-6,1) {Real};
    
    \node[inner sep=0pt] (fake) at (0,-1)
    {\incw{\pp-step1000}{2cm}\incw{\pp-step1000_y}{2cm}\incw{\pp-step1000_z}{2cm}~\incw{\pt-step1000}{2cm}\incw{\pt-step1000_y}{2cm}\incw{\pt-step1000_z}{2cm}};
    \node[inner sep=0pt] (target) at (-6,-1) {Proj.};
    
    \end{tikzpicture}
 
\caption{Results of projecting  real, unseen 3D images to the latent space. Top row shows the real image, and bottom row shows the reconstruction from the projected embedding. Configuration used was 2mm-fd96. The middle sagittal, axial and coronal slices of the 3D images are displayed. The differences between the real and reconstructed images are indistinguishable other than lattice-like artifacts caused by the generator.}
\label{fig:Projection}
\end{figure}
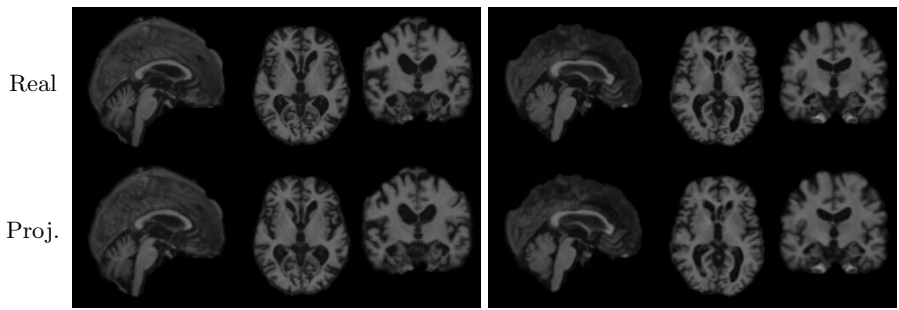

\noindent \textbf{3D Image Synthesis at 2mm Isotropic Resolution: }  
Figure~\ref{fig:GenImg_iso} shows randomly generated 2mm-resolution 3D images using 3D-StyleGAN that was trained with the configuration fd-96 with the filter map depth of 96 (see Table \ref{tab:config} for all configurations). Each model was trained for the average of $\sim$2 days.
We can observe that anatomical structures are generated correctly and coherently located.

In Table~\ref{tab:metrics}, we show the bMMD$^2$, MS-SSIM, and FIDs for the middle sagittal (FID-Sag), axial (FID-Ax) and coronal (FID-Cor) slices calculated using 4,000 generated images with respect to the unseen test images.
One interesting observation was that the slice-wise FIDs of middle sagittal and axial slices of the network trained with the filter map depth of 64 were lower than the one with the filter map depth of 96.
Figure~\ref{fig:96_64_comp} shows a qualitative comparison between images generated by networks with a filter map depth of 64 vs 96. Networks with a filter depth of 64 produce sharper boundaries in the cerebellum and sulci, but the overall brain structures are not as coherent as in the images with filter depth of 94. 
This is possibly caused by the characteristics of the FID score that focuses on the texture of the generated images, rather than taking into account the overall structures of the brain~\cite{zhang2018unreasonable}. 
The network with filter map depth of 32 showed the best bMMD$^2$ result while the distributions of the metrics were largely overlapped between fd-32, fd-64, and fd-96. The MS-SSIM showed that the network with filter mpa depth of 64 showed the most diversity in generated images. 
This may indicate that the better FID metrics of the fd-64 configuration was due to the diversity of generated images and the fd-96 configuration could be overfitted that did not cover the large variability of real images.

\begin{table}[tb!]
\centering
\caption{The quantitative evaluation results of generated images with different configurations. The Maximum Mean Discrepancy (bMMD$^2$) and Multi-Scale Structural Similarity Index (MS-SSIM) were calculated on each set of 3D generated and training real images. The slice-wise Fr\'{e}chet Inception Distance (FID) of the middle slices (Sag: Sagittal, Ax:Axial, and Cor:Coronal) of generated and training real images with respect to test unseen real images were summarized.}
\begin{tabular}{cccccc}
\Xhline{2\arrayrulewidth}
Configurations  & bMMD$^2$ & MS-SSIM & FID-Sag & FID-Ax & FID-Cor \\ \Xhline{2\arrayrulewidth}
2mm-fd16        &  7026 (555.79) &  0.96 (0.03)  & 145.6   & 153.6    & 164.1  \\ \hline
2mm-fd32        &  \textbf{4470} (532.65) &  0.94 (0.07)  & 129.3   & 144.3    & 128.8  \\ \hline
2mm-fd64        &  4497 (898.53) &  \textbf{0.93} (0.12)  & \textbf{106.9}   & \textbf{71.3}      & 90.2  \\ \hline
2mm-fd96        &  4475 (539.38) &  0.96 (0.04)  & 138.3   & 83.2      & \textbf{88.5}  \\ \hline 
2mm-Real        &  449 (121.43) &  0.85 (0.002)  & 3.0   & 2.1      & 2.9  \\ \Xhline{2\arrayrulewidth}
\end{tabular}
\label{tab:metrics}
\end{table}


\noindent \textbf{Image Projection: } 
Figure \ref{fig:Projection} shows the results of projecting unseen images into the space of latent vector $\mathbf{w}$ and noise map $\mathbf{n}$. Top row shows the real, unseen image, while the bottom row shows the reconstruction from the projected embedding. Reconstructed images are almost identical to the input images, a result that has also been observed by \cite{abdal2019image2stylegan} on other datasets, likely due to the over-parameterization of the latent space of StyleGAN2.

\begin{figure}[tb!]
\centering
  \subfloat[Sagittal View]{ \includegraphics[width=70mm]{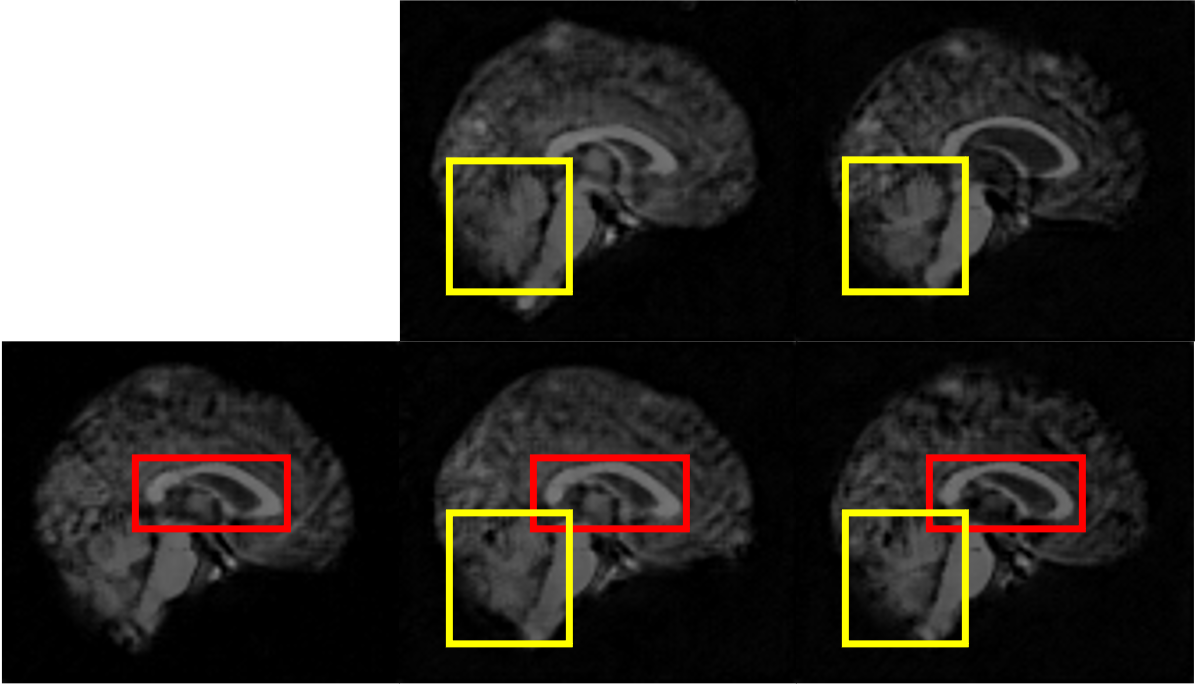} }
  \subfloat[Coronal View]{ \includegraphics[width=50mm]{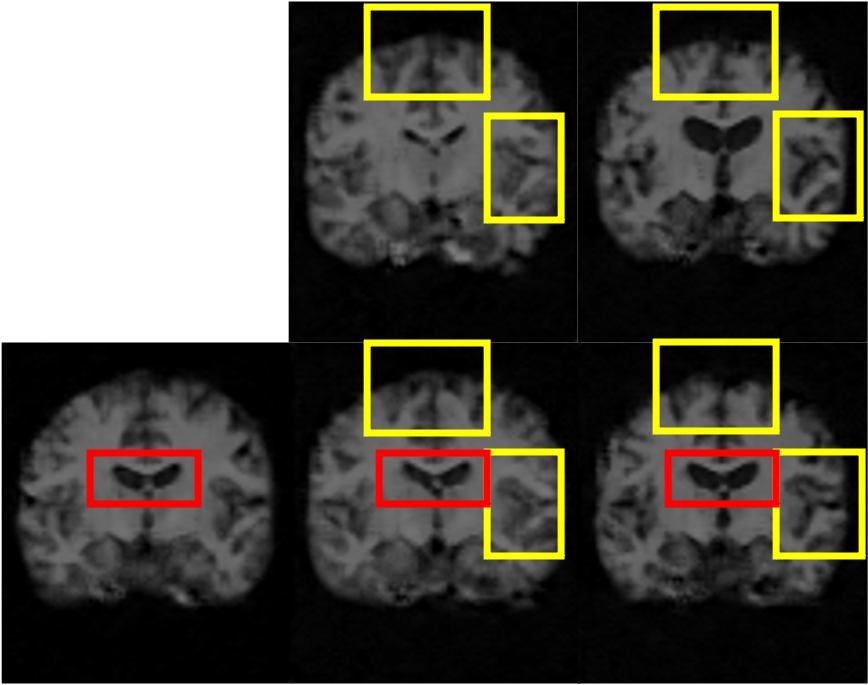} }
\caption{Style-mixing example. The style vectors of two 3D images at the top row at the high resolution (5\textsuperscript{th}-9\textsuperscript{th}) style blocks were mixed with the style vectors at at the low resolution (1\textsuperscript{st}-4\textsuperscript{th}) style blocks of an image at the first column were mixed and generated new images. Higher-level anatomical variations, such as the size of a ventricle and corpus callosum (red boxes), were controlled by the lower-resolution style blocks and the detailed variations, such as sulci and cerebellum structures (yellow boxes) by the higher-resolution style blocks. Three input images were randomly generated full 3D images and displayed on the (a) sagittal and (b) coronal views of the respective middle slices.}
\label{fig:StyleMixing}
\end{figure}

\noindent \textbf{Style Mixing: }Figure~\ref{fig:StyleMixing} shows the result of using 3D-StyleGAN's style mixing capability. The images at the top row and the first column are randomly generated from the trained 3D-StyleGAN with the configuration 2mm-fd96. The style vectors of the images at the top row were fed into the high resolution style blocks (i.e., the 5\textsuperscript{th} to 9\textsuperscript{th}). The style vector of the image at the first column is fed into the low resolution style blocks (i.e., the 1\textsuperscript{st} to 4\textsuperscript{th}).
We observed that high-level anatomical variations, such as the size of a ventricle and a brain outline, were controlled by the style vectors of the image at the first column mixed to the lower-resolution style blocks. 
These results indicate that the style vectors at different resolution levels could potentially control different levels of anatomical variability although our results are preliminary and need further investigation. 

\begin{figure}[htb!]
\centering
  \subfloat[1mm resolution]{ \includegraphics[width=60mm,height=20mm]{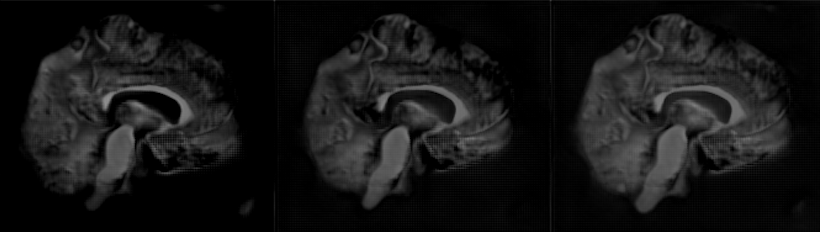} }
  \subfloat[2mm resolution]{ \includegraphics[width=60mm,height=20mm]{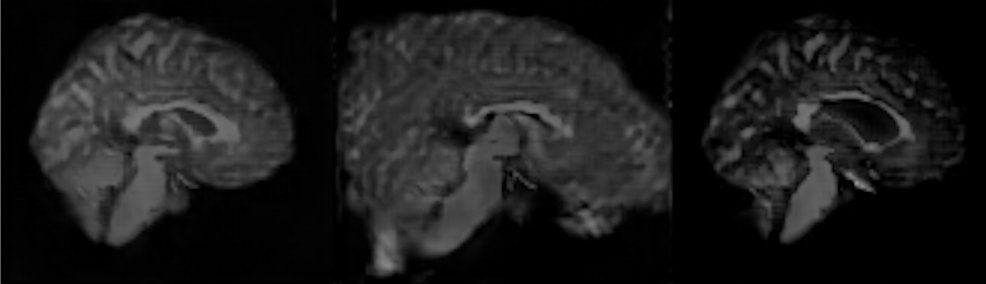} }
\caption{Failure cases of 1mm and 2mm isotropic resolution with the filter map depth of 16. The low filter map depth resulted in the low quality of images.}
\label{fig:Failure}
\end{figure}

\noindent \textbf{Failure Cases: } We trained the 3D-StyleGAN for the image synthesis of 1mm isotropic resolution images. The depths of filter map and latent vectors sizes were set to 16 and 32, respectively, because of the limited GPU memory (12GB). The model was trained for $\sim$5 days. Figure~\ref{fig:Failure} (a) shows the examples of generated 1mm-resolution images on the sagittal view after 320 iterations. The results showed the substantially lower quality of generated images compared to those from the networks trained with deeper filter depths for 2mm-resolution images.
Figure~\ref{fig:Failure} (b) shows the image generation result for 2mm-resolution images with the same filter depth, 16. Compared to the generated images shown in Figure~\ref{fig:GenImg_iso}, the quality of generated images is also substantially lower. 
This experiment showed that the filter map depth of generator and discriminator networks need to be above certain thresholds to assure the high quality of generated images.

\section{Discussion}
\label{sec:Discussion}

We presented 3D-StyleGAN, an extension of StyleGAN2 for the generative modeling of 3D medical images. 
In addition to the high quality of generated images of 3D-StyleGAN, we believe the latent space projection of unseen real images and the control of anatomical variability through style vectors can be utilized to tackle clinically important problems.
One limitation of our paper is in the use of limited evaluation metrics. The evaluation metrics did not match with qualitative evaluations on the generated image qualities.
We used FIDs that only worked on 2D slices due to their reliance on the 2D VGG network. We plan to address 3D quantitative evaluation metrics on perceptual similarity with a large 3D network in future work.
Another limitation of our work is the limited sizes of the training data and the trained networks. We plan to address this in future work through extensive augmentation of the training images and investigating memory-efficient network structures. We believe our work will enable important downstream tasks to be done directly in 3D via generative modeling, such as super-resolution, motion correction, conditional atlas estimation, and image progression modeling with respect to clinical attributes.

\vspace{10pt}

\noindent \textbf{Acknowledgements} This research was supported by NIH-NINDS MRI-GENIE: R01NS086905; K23NS064052, R01NS082285, NIH NIBIB NAC P41EB015902, NIH NINDS U19NS115388, Wistron, and MIT-IBM Watson AI Lab. A.K.B. was supported by MGH ECOR Fund for Medical Discovery (FMD) Clinical Research Fellowship Award. M.B. was supported by the ISITE and Planiol foundations and the Sociétés Françaises de Neuroradiologie et Radiologie.


\bibliographystyle{splncs03}

\begin{thebibliography}{10}
\providecommand{\url}[1]{\texttt{#1}}
\providecommand{\urlprefix}{URL }

\bibitem{tensorflow2015-whitepaper}
Abadi, M., Agarwal, A., Barham, P., Brevdo, E., Chen, Z., Citro, C., Corrado,
  G.S., Davis, A., Dean, J., Devin, M., Ghemawat, S., Goodfellow, I., Harp, A.,
  Irving, G., Isard, M., Jia, Y., Jozefowicz, R., Kaiser, L., Kudlur, M.,
  Levenberg, J., Man\'{e}, D., Monga, R., Moore, S., Murray, D., Olah, C.,
  Schuster, M., Shlens, J., Steiner, B., Sutskever, I., Talwar, K., Tucker, P.,
  Vanhoucke, V., Vasudevan, V., Vi\'{e}gas, F., Vinyals, O., Warden, P.,
  Wattenberg, M., Wicke, M., Yu, Y., Zheng, X.: {TensorFlow}: Large-scale
  machine learning on heterogeneous systems (2015),
  \url{https://www.tensorflow.org/}, software available from tensorflow.org

\bibitem{abdal2019image2stylegan}
Abdal, R., Qin, Y., Wonka, P.: Image2stylegan: How to embed images into the
  stylegan latent space? In: Proceedings of the IEEE/CVF International
  Conference on Computer Vision. pp. 4432--4441 (2019)

\bibitem{arjovsky2017wasserstein}
Arjovsky, M., Chintala, S., Bottou, L.: Wasserstein generative adversarial
  networks. In: International conference on machine learning. pp. 214--223.
  PMLR (2017)

\bibitem{brock2018large}
Brock, A., Donahue, J., Simonyan, K.: Large scale gan training for high
  fidelity natural image synthesis. arXiv preprint arXiv:1809.11096  (2018)

\bibitem{dagley2017harvard}
Dagley, A., LaPoint, M., Huijbers, W., Hedden, T., McLaren, D.G., Chatwal,
  J.P., Papp, K.V., Amariglio, R.E., Blacker, D., Rentz, D.M., et~al.: Harvard
  aging brain study: dataset and accessibility. Neuroimage  144,  255--258
  (2017)

\bibitem{dalca2018anatomical}
Dalca, A.V., Guttag, J., Sabuncu, M.R.: Anatomical priors in convolutional
  networks for unsupervised biomedical segmentation. In: Proceedings of the
  IEEE Conference on Computer Vision and Pattern Recognition. pp. 9290--9299
  (2018)

\bibitem{di2014autism}
Di~Martino, A., Yan, C.G., Li, Q., Denio, E., Castellanos, F.X., Alaerts, K.,
  Anderson, J.S., Assaf, M., Bookheimer, S.Y., Dapretto, M., et~al.: The autism
  brain imaging data exchange: towards a large-scale evaluation of the
  intrinsic brain architecture in autism. Molecular psychiatry  19(6),
  659--667 (2014)

\bibitem{dumoulin2018feature}
Dumoulin, V., Perez, E., Schucher, N., Strub, F., Vries, H.d., Courville, A.,
  Bengio, Y.: Feature-wise transformations. Distill  3(7),  e11 (2018)

\bibitem{fischl2012freesurfer}
Fischl, B.: Freesurfer. Neuroimage  62(2),  774--781 (2012)

\bibitem{gollub2013mcic}
Gollub, R.L., Shoemaker, J.M., King, M.D., White, T., Ehrlich, S., Sponheim,
  S.R., Clark, V.P., Turner, J.A., Mueller, B.A., Magnotta, V., et~al.: The
  mcic collection: a shared repository of multi-modal, multi-site brain image
  data from a clinical investigation of schizophrenia. Neuroinformatics  11(3),
   367--388 (2013)

\bibitem{goodfellow2014generative}
Goodfellow, I.J., Pouget-Abadie, J., Mirza, M., Xu, B., Warde-Farley, D.,
  Ozair, S., Courville, A., Bengio, Y.: Generative adversarial networks. arXiv
  preprint arXiv:1406.2661  (2014)

\bibitem{gretton2012kernel}
Gretton, A., Borgwardt, K.M., Rasch, M.J., Sch{\"o}lkopf, B., Smola, A.: A
  kernel two-sample test. The Journal of Machine Learning Research  13(1),
  723--773 (2012)

\bibitem{hara3dcnns}
Hara, K., Kataoka, H., Satoh, Y.: Can spatiotemporal 3d cnns retrace the
  history of 2d cnns and imagenet? In: Proceedings of the IEEE Conference on
  Computer Vision and Pattern Recognition (CVPR). pp. 6546--6555 (2018)

\bibitem{he2016deep}
He, K., Zhang, X., Ren, S., Sun, J.: Deep residual learning for image
  recognition. In: Proceedings of the IEEE conference on computer vision and
  pattern recognition. pp. 770--778 (2016)

\bibitem{heusel2017gans}
Heusel, M., Ramsauer, H., Unterthiner, T., Nessler, B., Hochreiter, S.: Gans
  trained by a two time-scale update rule converge to a local nash equilibrium.
  arXiv preprint arXiv:1706.08500  (2017)

\bibitem{holmes2015brain}
Holmes, A.J., Hollinshead, M.O., O’keefe, T.M., Petrov, V.I., Fariello, G.R.,
  Wald, L.L., Fischl, B., Rosen, B.R., Mair, R.W., Roffman, J.L., et~al.: Brain
  genomics superstruct project initial data release with structural,
  functional, and behavioral measures. Scientific data  2(1),  1--16 (2015)

\bibitem{huang2017arbitrary}
Huang, X., Belongie, S.: Arbitrary style transfer in real-time with adaptive
  instance normalization. In: Proceedings of the IEEE International Conference
  on Computer Vision. pp. 1501--1510 (2017)

\bibitem{karras2017progressive}
Karras, T., Aila, T., Laine, S., Lehtinen, J.: Progressive growing of gans for
  improved quality, stability, and variation. arXiv preprint arXiv:1710.10196
  (2017)

\bibitem{karras2020training}
Karras, T., Aittala, M., Hellsten, J., Laine, S., Lehtinen, J., Aila, T.:
  Training generative adversarial networks with limited data. arXiv preprint
  arXiv:2006.06676  (2020)

\bibitem{karras2019style}
Karras, T., Laine, S., Aila, T.: A style-based generator architecture for
  generative adversarial networks. In: Proceedings of the IEEE/CVF Conference
  on Computer Vision and Pattern Recognition. pp. 4401--4410 (2019)

\bibitem{karras2020analyzing}
Karras, T., Laine, S., Aittala, M., Hellsten, J., Lehtinen, J., Aila, T.:
  Analyzing and improving the image quality of stylegan. In: Proceedings of the
  IEEE/CVF Conference on Computer Vision and Pattern Recognition. pp.
  8110--8119 (2020)

\bibitem{kwon2019generation}
Kwon, G., Han, C., Kim, D.s.: Generation of 3d brain mri using auto-encoding
  generative adversarial networks. In: International Conference on Medical
  Image Computing and Computer-Assisted Intervention. pp. 118--126. Springer
  (2019)

\bibitem{maas2013rectifier}
Maas, A.L., Hannun, A.Y., Ng, A.Y.: Rectifier nonlinearities improve neural
  network acoustic models. In: Proc. icml. vol.~30, p.~3. Citeseer (2013)

\bibitem{marcus2007open}
Marcus, D.S., Wang, T.H., Parker, J., Csernansky, J.G., Morris, J.C., Buckner,
  R.L.: Open access series of imaging studies (oasis): cross-sectional mri data
  in young, middle aged, nondemented, and demented older adults. Journal of
  cognitive neuroscience  19(9),  1498--1507 (2007)

\bibitem{marek2011parkinson}
Marek, K., Jennings, D., Lasch, S., Siderowf, A., Tanner, C., Simuni, T.,
  Coffey, C., Kieburtz, K., Flagg, E., Chowdhury, S., et~al.: The parkinson
  progression marker initiative (ppmi). Progress in neurobiology  95(4),
  629--635 (2011)

\bibitem{milham2012adhd}
Milham, M.P., Fair, D., Mennes, M., Mostofsky, S.H., et~al.: The adhd-200
  consortium: a model to advance the translational potential of neuroimaging in
  clinical neuroscience. Frontiers in systems neuroscience  6, ~62 (2012)

\bibitem{mueller2005ways}
Mueller, S.G., Weiner, M.W., Thal, L.J., Petersen, R.C., Jack, C.R., Jagust,
  W., Trojanowski, J.Q., Toga, A.W., Beckett, L.: Ways toward an early
  diagnosis in alzheimer’s disease: the alzheimer’s disease neuroimaging
  initiative (adni). Alzheimer's \& Dementia  1(1),  55--66 (2005)

\bibitem{odena2017conditional}
Odena, A., Olah, C., Shlens, J.: Conditional image synthesis with auxiliary
  classifier gans. In: International conference on machine learning. pp.
  2642--2651. PMLR (2017)

\bibitem{simonyan2014very}
Simonyan, K., Zisserman, A.: Very deep convolutional networks for large-scale
  image recognition. arXiv preprint arXiv:1409.1556  (2014)

\bibitem{volokitin2020modelling}
Volokitin, A., Erdil, E., Karani, N., Tezcan, K.C., Chen, X., Van~Gool, L.,
  Konukoglu, E.: Modelling the distribution of 3d brain mri using a 2d slice
  vae. In: International Conference on Medical Image Computing and
  Computer-Assisted Intervention. pp. 657--666. Springer (2020)

\bibitem{zhang2018unreasonable}
Zhang, R., Isola, P., Efros, A.A., Shechtman, E., Wang, O.: The unreasonable
  effectiveness of deep features as a perceptual metric. In: Proceedings of the
  IEEE conference on computer vision and pattern recognition. pp. 586--595
  (2018)

\bibitem{zhu2017unpaired}
Zhu, J.Y., Park, T., Isola, P., Efros, A.A.: Unpaired image-to-image
  translation using cycle-consistent adversarial networks. In: Proceedings of
  the IEEE international conference on computer vision. pp. 2223--2232 (2017)

\end{thebibliography}

\end{document}